\documentstyle[epsfig,12pt]{article}
\begin{document}
\baselineskip=15pt

\begin{center}
{\Huge \bf --------------------------}

{\tiny

SUNY Institute of Technology at Utica/Rome\\

Fall 2001 Conference on Theoretical High Energy Physics\\

 }


{\Huge \bf --------------------------}

{\Huge \bf --------------------------}

\end{center}

{\small \sl

}

\sf

\centerline{\large\bf CHIRAL SYMMETRY ON ${\bf S}^2_F$}
\bigskip
\begin{center}
Giorgio Immirzi$^{*}$,
Badis Ydri$^{+}$\footnote{Previous Address : Department of Physics , Syracuse University , Syracuse , NY 13244-1130 , U.S.A .}\\
\bigskip
$^{*}${\it Universita' di Perugia and INFN,
Perugia , Italy .}\\
\bigskip
$^{+}${\it School of Theoretical Physics , Dublin Institute for Advanced Studies ,\\
10 Burlington Road , Dublin 4 , Ireland .}\\

\vskip.5cm

\begin{abstract}
In this talk we give a brief description of the formulation of
chiral and gauge symmetries on the  fuzzy sphere . In particular
fermion doublers are shown to be absent and the correct anomaly
equation in two dimensions is obtained in the corresponding
continuum limit .
\end{abstract}

\end{center}

A fuzzy space is by construction a discrete lattice-like
structure which serves to regularize \cite{ydri}, it allows for
an exact chiral invariance to be formulated , but still the
fermion-doubling problem is completely avoided \cite{trg}. Global
chiral anomaly was also , along with other topological
non-trivial field configurations , formulated in
\cite{bal,grosse} .

In this talk we sketch how we can go beyond global considerations
and define a "fuzzy" chiral anomaly associated with a "fuzzy"
$U(1)$ global chiral symmetry . This "fuzzy" chiral anomaly has
the correct continuum limit in the sense that it approaches , in
the limit of large IRR's $l$ of $SU(2)$ , the "local" axial
anomlay on continuum ${\bf S}^2$ with a corresponding canonical
theta term .

\section{Continuum Action}

The free fermionic action on continuum ${\bf S^2}$ is given by
\cite{presnajder,denjoe}
\begin{eqnarray}
S=\int_{{\bf S}^2} \frac{d{\Omega}}{4{\pi}}\sum_{{\alpha}{\beta}}
\bar{\chi}^{\alpha}{\cal
D}^{{\alpha}{\beta}}{\chi}^{\beta},\label{contaction0}
\end{eqnarray}
where ${\chi}^{\alpha}$ , ${\alpha}=1,2$ , are the two components
of the continuum spinor which are both smooth functions on ${\bf
S}^2$ . ${\cal D}$ is the Dirac operator on ${\bf S}^2$ given by
\cite{ydri,trg,bal,grosse}
\begin{eqnarray}
{\cal D}&=&{\cal P}_{ij}{\sigma}_i({\cal
L}_j+\frac{1}{2}{{\sigma}}_j){\equiv}\vec{\sigma}\vec{\cal
L}+1,~{\cal
L}_k=-i{\epsilon}_{kij}n_i{\partial}_j.\label{contdirac}
\end{eqnarray}
This Dirac operator admits the chirality operator $
{\gamma}=\vec{\sigma}.\vec{n}$ ,
${\gamma}^2=1~,~{\gamma}^{+}={\gamma}~$ and such that
$\{{\gamma},{\cal D}\}=0$ .

The projector ${\cal P}$ which defines the tangent bundle ${\bf
T}{\bf S}^2$ of ${\bf S}^2$ is given by $ {\cal
P}=(\vec{n}.\vec{\theta})^2~,~{\rm
where}~({\theta}_i)_{jk}=-i{\epsilon}_{ijk}$ , and satisfies
${\cal P}^2={\cal P}$ and ${\cal P}^{+}={\cal P}$ . So given any
vector $\vec{A}$ in ${\bf R}^3$ , ${\cal P}\vec{A}$ defines a
vector tangent to ${\bf S}^2$ . Indeed ${\cal P}\vec{A}$ has only
two independent components tangent to ${\bf S}^2$ as one can
check if we rewrite it in terms of its components, namely ${\cal
P}_{ij}={\delta}_{ij}-n_in_j$ , and then compute that
$\vec{n}.{\cal P} \vec{A}=n_i{\cal P}_{ij}{A}_j=0$ .

Gauging the Dirac operator (\ref{contdirac}) means invoking the
minimal replacement ${\cal L}_i{\longrightarrow}{\nabla}_{i}={\cal
L}_i+{A}_i$ , where $\vec{A}$ is a gauge field which is generally
in ${\bf R}^3$ satisfying ${A}^{+}_i={A}_i$. The gauged Dirac
operator is therefore
\begin{eqnarray}
{\cal D}_{G}={\cal P}_{ij}{\sigma}_i({\nabla}_{j}+
\frac{1}{2}{{\sigma}}_j)\equiv\vec{\sigma}\vec{\cal L}
+1+{{\sigma}}_i\hat{A}_i.\label{contgauge}
\end{eqnarray}
The gauge connection on ${\bf S}^2$ is defined by
$\vec{\hat{A}}={\cal P}\vec{A}$ with components $\hat{A}_i={\cal
P}_{ij}{A}_j $ , and correspondingly , the gauged fermionic
action is given by
\begin{eqnarray}
S_{G}&=&\int_{{\bf
S}^2}\frac{d{\Omega}}{4{\pi}}\sum_{\alpha\beta}\bar{\chi}_{\alpha}{\cal
D}_{G}^{\alpha\beta}{\chi}^{\beta},\label{contaction1}
\end{eqnarray}
The corresponding pure $U(1)$ Yang-Mills action is of the form
\begin{equation}
S_{YM}=-\frac{1}{4e^2}\int_{{\bf S}^2}
\frac{d{\Omega}}{4{\pi}}{F}_{ij}{F}_{ij},
\end{equation}
where the $U(1)$ curvature ${F}_{ij}$ is defined by
\begin{eqnarray}
{F}_{ij}&=&[{\nabla}_i,{\nabla}_{j}]-i{\epsilon}_{ijk}{\nabla}_{k}{\equiv}[{\cal
L}_i,{{A}}_j]-[{\cal L}_j,{{A}}_i]-i{\epsilon}_{ijk}A_k.
\end{eqnarray}
$U(1)$ gauge transformations ,
$U=U(\vec{n})=e^{i{\Lambda}(\vec{n})}$ , act on the spinor
${\chi}$ , on the covariant derivative ${\nabla}_{i}$ and on the
curvature ${F}_{ij}$ in the usual fashion
\begin{eqnarray}
&&{\chi}{\longrightarrow}{\chi}^{'}=U{\chi}\nonumber\\
&&{\nabla}_{i}{\longrightarrow}{\nabla}_{i}^{'}=U{\nabla}_{i}U^{+}\nonumber\\
&&~{F}_{ij}{\longrightarrow}{F}_{ij}^{'}={F}_{ij}.
\end{eqnarray}

\section{Regularization via Fuzzification}

Instead of replacing Euclidean space-time ${\bf S}^2$ by a lattice
, we will implement in the following the regularization
prescription given by the substitution ${\bf
S}^2{\longrightarrow}{\bf S}^2_F$ where ${\bf S}^2_F$ is the fuzzy
sphere [ see \cite{ydri} and references therein ] . We first make
the following replacement
\begin{equation}
n_i{\longrightarrow}n_i^F=\frac{L_i}{\sqrt{l(l+1)}},\label{2}
\end{equation}
where $L_i$'s are the generators  of the IRR $l$ of $SU(2)$ , i.e
they satisfy $\sum_{i=1}^3L_i^2=l(l+1)$ and
$[L_i,L_j]=i{\epsilon}_{ijk}L_k$ . Loosely speaking , ${\bf
S}^2_F$ is the algebra ${\bf A}$ of all $(2l+1){\times}(2l+1)$
matrices which is generated by the $L_i$'s , i.e ${\bf
A}=Mat_{2l+1}$.

At the level of the action we also make the following replacements
\begin{eqnarray}
&&\int_{{\bf S}^2}\frac{d{\Omega}}{4{\pi}}{\longrightarrow} \frac{1}{2(2l+1)}Tr_{H_l^{(2)}}\nonumber\\
&&{\cal D}=\vec{\sigma}\vec{\cal L}+1{\longrightarrow}{D}_{F}=\vec{\sigma}[\vec{L},.]+1\nonumber\\
&&{\chi}^{\alpha}{\longrightarrow}{\psi}_F^{\alpha}\nonumber\\
&&{A}_i{\longrightarrow}A_i^F.
\end{eqnarray}
$H_{l}^{(2)}$ is the Hilbert space generated by the $SU(2)$
coherent states $|\vec{n},l>$ \cite{lee} . In particular we have the identity
\begin{eqnarray}
\frac{1}{2l+1}Tr_{H_l^{(2)}}X=\int \frac{d\Omega}{4{\pi}}<\vec{n},l|X|\vec{n},l>.
\end{eqnarray}
${\psi}_F^{\alpha}$ ,
${\alpha}=1,2$ , as well as $A_i^F$ , $i=1,2,3$ , are now
$(2l+1){\times}(2l+1)$ matrices. The fuzzy action is therefore
\begin{eqnarray}
&&S_{GF}=\frac{1}{2(2l+1)}Tr_{H_l^{(2)}}\sum_{\alpha \beta}
\bar{{\psi}}_F^{\alpha}{D}_{GF}^{\alpha \beta}
{\psi}_F^{\beta}\nonumber\\
&&{D}_{GF}={D}_F+{\sigma}_iA_i^F. \label{fuzzyaction1}
\end{eqnarray}
${D}_F$ above is the Grosse-Klim\v{c}\'{i}k-Pre\v{s}najder Dirac
operator on fuzzy ${\bf S}^2$ , it can be rewritten in the form
${D}_{ F}=\vec{\sigma}.\vec{L}^L-\vec{\sigma}.\vec{L}^R+1$ ,
where $L_i^L$'s and $-L_i^R$'s are the generators of the IRR $l$
of $SU(2)$ which act respectively on the left and on the right of
the algebra ${\bf A}$ \cite{ydri,trg, bal,grosse}. This Dirac
operator has the correct continuum limit in the sense  that
${D}_F{\longrightarrow}{\cal D}$.

The chirality operator on ${\bf S}^2_F$ was , on the other hand ,
first found in \cite{watamuras}, it is given by
\begin{eqnarray}
{\Gamma}^R&=&\frac{1}{l+\frac{1}{2}}[-\vec{\sigma}.\vec{L}^R+\frac{1}{2}],\label{chiralities}
\end{eqnarray}
which also satisfies
\begin{eqnarray}
{\Gamma}^R{D}_{F}+{D}_{F}{\Gamma}^R&=&\frac{1}{l+\frac{1}{2}}{D}^2_{F}.\label{GWrelation}
\end{eqnarray}
Despite the fact that this anticommutation relation is not exact ,
one can show that $({D}_{F},{\Gamma}^R)$ defines a chiral
structure on fuzzy ${\bf S}^2$ which satisfies $a)$ the
Ginsparg-Wilson relation , $b)$ is without fermion doubling and
$c)$ has the correct continuum limit \cite{ydri,trg}.

The absence of fermion doubling can be easily seen from comparing
the spectrum of ${D}_{F}$ which is given by \cite{grosse}
\begin{eqnarray}
{D}_{F}(j)&=&{\pm}(j+\frac{1}{2}) , j=\frac{1}{2},\frac{3}{2},...,2l-\frac{1}{2}\nonumber\\
&=&j+\frac{1}{2}~~ ~{\rm for} ~~~j=2l+\frac{1}{2},\label{fuzzyspectrum}
\end{eqnarray}
with the spectrum of ${\cal D}$ given by
\begin{eqnarray}
{\cal D}(j)&=&{\pm}(j+\frac{1}{2}) ,
j=\frac{1}{2},\frac{3}{2},...,\infty.
\end{eqnarray}
As one can immediately see there is no fermion doubling and the
spectrum of ${D}_F$ is simply cut-off at the top eigenvalue
$j=2l+\frac{1}{2}$ if compared to the continuum spectrum
\cite{ydri,trg}.

Similarly to the continuum case , the fuzzy gauge potential
$A_i^F$ must be projected onto the sphere in order to have a
fuzzy gauge theory {\it localized} in some appropriate sense on
${\bf S}^2_F$ . The simplest way to this end is to construct the
fuzzy analogue $P$ of the continuum projector ${\cal P}$ which
will define the fuzzy tangent bundle ${\bf T}{\bf S}^2_F$ of
${\bf S}^2_F$. It can be derived without any difficulty and it
turns out to be given by
\begin{eqnarray}
P_{ij}={\delta}_{ij}-n_i^Fn_j^F
\end{eqnarray}
The vector $P_{ij}A_j^F$ is indeed normal to the fuzzy sphere in
the sense that we have
${n}^F_i.{\hat{A}}^F_i=n_i^FP_{ij}{A}_j^F=0$ . Beside the fact
that the components $\hat{A}_i^F$'s are not self-adjoint , the
transversality condition  ${n}^F_i.{\hat{A}}^F_i=0$ is not stable
under fuzzy gauge transformations , and this , in turn ,
introduces a lot of complications into the problem which are
addressed in great detail in \cite{badisgiorgio} . In particular we showed there that the normal component of the fuzzy gauge potential can be
projected out by making it infinitely heavy through a gauge-invariant mass term in the action , and hence it effectively decouples .

Finally , and as was shown in \cite{badisgiorgio} , a fuzzy gauge
principle can also be written down , under which the fuzzy gauged
fermionic action (\ref{fuzzyaction1}) will indeed be invariant .
This fuzzy $U(1)$ gauge theory will have as a continuum limit the
ordinary $U(1)$ gauge theory defined earlier in section $1$ .

\section{Chiral Symmetry and Axial Anomaly}

On continuum ${\bf S}^2$ exact chiral invariance of the classical
action is expressed by the anti-commutation relation
\begin{eqnarray}
{\gamma}{\cal D}+{\cal D}{\gamma}=0.\label{contsense}
\end{eqnarray}
It is a well known fact that this symmetry will be broken by
quantum effects , and different methods of regularization are
shown to give rise to the same topological action . We showed in
\cite{badisgiorgio} that the fuzzy sphere is a novel
regularization scheme which leads also to the canonical theta
term. This result will now be skteched .

In spite of the Ginsparg-Wilson relation (\ref{GWrelation}) ,
exact chiral invariance on fuzzy ${\bf S}^2$ in the sense of
(\ref{contsense}) can also be constructed as was shown first in
\cite{trg}. Following \cite{badisgiorgio} we adopt here a
different route in defining chiral invariance . First we start by
rewriting the Ginsparg-Wilson relation (\ref{GWrelation}) in the
suggestive form \cite{leonardo}
\begin{eqnarray}
-{\Gamma}^L{D}_F+{D}_F{\Gamma}^R=0,\label{GW2}
\end{eqnarray}
where ${\Gamma}^R$ is given by equation (\ref{chiralities}) , and
${\Gamma}^L$ is obtained from (\ref{chiralities}) through the
substitution $-L^R_i{\longrightarrow}{L}^L_i$ . As in the
continuum , {\it Fuzzy chiral transformations} will be defined by
\begin{eqnarray}
&&{\psi}_F{\longrightarrow}{\psi}^{'}_F={\psi}_F+{\delta}{\psi}_F~,~{\delta}{\psi}_F={\Gamma}^{R}{\psi}_F{\lambda}^L\nonumber\\
&&\bar{\psi}_F{\longrightarrow}\bar{\psi}^{'}_F=\bar{\psi}_F+{\delta}\bar{\psi}_F~,~{\delta}\bar{\psi}_F=-{\lambda}^L\bar{\psi}_F{\Gamma}^L.\label{fuzzytrans2}
\end{eqnarray}
These transformations , as it turns out , do not leave  the
action (\ref{fuzzyaction1}) invariant but instead leave invariant
the action
\begin{eqnarray}
S_{CF}&=&\frac{1}{2(2l+1)}Tr_{H_{l}^{(2)}}\sum_{{\alpha}{\beta}}\bigg[\bar{\psi}_{F}^{\alpha}D_{CF}^{{\alpha}{\beta}}{\psi}_F^{\beta}\bigg]\nonumber\\
D_{CF}&=&D_F+{\epsilon}_{ijk}Z_j^Fn^F_k{A}_{i}^F.\label{fuzzyaction2}
\end{eqnarray}
Indeed , the change of this action under these transformations is
given by
\begin{eqnarray}
{\Delta}S_{CF}&=&-\frac{1}{2(2l+1)}Tr_{H_l^{(2)}}{\lambda}^L[L_i,\bar{{\psi}_F}{\sigma}_i{\Gamma}^R{{\psi}_F}].\label{divergenceF}
\end{eqnarray}
The connection between (\ref{fuzzyaction1}) and
(\ref{fuzzyaction2}) is discussed thoroughly in
\cite{badisgiorgio} . In particular $Z_k^F$ is given by $
Z_k^F=\frac{i}{2}[{\Gamma}^L{\sigma}_k+{\sigma}_k{\Gamma}^R]$ and
hence both actions tend in the large $l$ limit to the same
continuum action (\ref{contaction1}) . ${\lambda}^L$ in all the
above equations is a test $(2l+1){\times}(2l+1)$ matrix which is
infinitesimal in some appropriate sense \cite{badisgiorgio}.

The continuum limit of (\ref{divergenceF}) was computed in
\cite{badisgiorgio} and is given by
\begin{eqnarray}
{\Delta}{S}_{CF}{\longrightarrow}{\Delta}{S}_{C}=\frac{2lnl}{(4{\pi})^5}\int_{{\bf
S}^2}\frac{d{\Omega}}{4{\pi}}{\lambda}(\vec{n}){\cal
L}_i\bigg[\bar{\chi}{\sigma}_i{\gamma}{\chi}\bigg](\vec{n}).\label{anomaly0}
\end{eqnarray}

\subsection{The Theta Term}

Since we are dealing with a matrix model , manipulations on the
quantum measure will all have a well defined meaning . Following
\cite{fujikawa} we first expand the fuzzy spinors ${\psi}_F$ and
$\bar{\psi}_F$ in terms of the eigen-tensors ${\psi}_F(a,A)$ of
the gauged Dirac operator ${D}_{GF}$ as follows
\begin{eqnarray}
{\psi}_F=\sum_{a}{\theta}_{a}{\psi}_F(a,A)~,~\bar{\psi}_F=\sum_{a}\bar{\theta}_{a}{\psi}_{F}^{+}(a,A),\label{expansion}
\end{eqnarray}
where ${\theta}_a$'s and $\bar{\theta}_a$'s are two independent sets of Grassmanian variables , and
${\psi}_F(a,A)$'s are defined by
\begin{eqnarray}
{D}_{GF}{\psi}_F(a,A)=e_a(A){\psi}_F(a,A),
\end{eqnarray}
and are normalized such that
\begin{eqnarray}
\frac{1}{2(2l+1)}Tr_{H_l^{(2)}}{\psi}_F^{+}(a,A){\psi}_{F}(b,A)={\delta}_{ab}.\label{normalization}
\end{eqnarray}
$a$ stands for all the quantum numbers needed to characterize the
eigenvalues of the Dirac operator ${D}_{GF}$ . For weak fuzzy
gauge fields $A_i^F$'s, $a$ stands for $j$ , $k$ and $m$ which
are the eigenvalues of
$\vec{J}^2=(\vec{K}+\frac{\vec{\sigma}}{2})^2$ ,
$\vec{K}^2=(\vec{L}^L-\vec{L}^R)^2$ and $J_3$ respectively .

As it turns out \cite{badisgiorgio} , and similarly to \cite{fujikawa} , what really matters in the calculation is the
asymptotic behaviour of ${\psi}_F(a,A)$'s and $e_a(A)$'s given by \cite{ydri}
\begin{eqnarray}
&&e_{a}(A)_{A^F_i{\longrightarrow}0}{\longrightarrow}e_{a}=j(j+1)-k(k+1)+\frac{1}{4}\nonumber\\
&&{\psi}_{F}(a,A)_{A^F_i{\longrightarrow}0}{\longrightarrow}{\psi}_{F}(a)=\sqrt{2(2l+1)}\sum_{k_3,\sigma}C^{jm}_{kk_3\frac{1}{2}\sigma}T_{kk_3}(l)
{\chi}_{\frac{1}{2}\sigma}.
\end{eqnarray}
The sum over $a$ in (\ref{expansion}) is therefore finite and given by
\begin{eqnarray}
\sum_{a}=\sum_{k=0}^{2l}\sum_{j=k-\frac{1}{2}}^{k+\frac{1}{2}}\sum_{m=-j}^{j},
\end{eqnarray}
and hence the quantum measure is also well defined
\begin{eqnarray}
{\cal D}{\psi}_F{\cal
D}\bar{\psi}_F=\prod_{a}d{\theta}_ad{\bar{\theta}_a}{\longrightarrow}\prod_{k=0}^{2l}\prod_{j=k-\frac{1}{2}}^{k+\frac{1}{2}}\prod_{m=-j}^{j}d{\theta}_{kjm}d{\bar{\theta}_{kjm}},
\end{eqnarray}
A canonical calculation shows that the above quantum measure
changes under the fuzzy chiral transformations
(\ref{fuzzytrans2}) as follows
\begin{eqnarray}
\int {\cal D}{{\psi}_F}^{'}{\cal
D}\bar{{\psi}}_F^{'}e^{-S^{'}_{CF}}=\int {\cal D}{{\psi}_F}{\cal
D}\bar{{\psi}_F}e^{S_{{\theta}F}}e^{-S_{CF}-{\Delta}S_{CF}}.\label{partition}
\end{eqnarray}
The theta term on ${\bf S}^2_F$ is therefore given by
\begin{eqnarray}
S_{{\theta}F}&=&-\frac{1}{2(2l+1)}\sum_{a}Tr_{H_{l}^{(2)}}{\lambda}^L{\psi}_F^{+}(a,A)({\Gamma}^R-{\Gamma}^L){\psi}_F(a,A).\label{anomaly}\nonumber\\
\end{eqnarray}
We will now extract the large $l$ behaviour of this last formula
, i.e we will carefully analyze the spectrum of the theory and
derive the continuum limit of the axial anomaly . We first start
with the free theory and rewrite the Ginsparg-Wison relation
(\ref{GW2}) in the form
\begin{eqnarray}
({\Gamma}^R-{\Gamma}^L)D_F+D_F({\Gamma}^R-{\Gamma}^L)=0,\label{importantfact}
\end{eqnarray}
which means that in the absence of gauge field we must have
\begin{eqnarray}
tr[{\Gamma}^R-{\Gamma}^L]=0.
\end{eqnarray}
The trace is meant to be in the space of spinors . This is also
true in the {\it naive } continuum free theory , i.e
${\gamma}{\cal D}+{\cal D}{\gamma}=0$ , $tr{\gamma}=0$ . As we
will show and if we have to be precise this result is only valid
for the infrared sector of the theory . However if we include the
gauge field we can compute instead that
\begin{eqnarray}
({\Gamma}^R-{\Gamma}^L)D_{GF}+D_{GF}({\Gamma}^R-{\Gamma}^L)&=&\{{\Gamma}^R-{\Gamma}^L,{\sigma}_iA_i^F\}.
\label{ginsparg-wilson}
\end{eqnarray}
The naive limit of this equation is ${\gamma}{\cal D}_G+{\cal
D}_G{\gamma}=2{\phi}$ where ${\phi}$ is the normal component of
the gauge field , namely ${\phi}=\vec{n}.\vec{A}$ , and hence it
will be eventually set equal to zero by means of the projector
${\cal P}$ for example [ see \cite{badisgiorgio} for a different
method of projecting the gauge potential onto the sphere ]. In
other words , in the continuum interacting theory one might be
tempted to conclude that $Tr{\gamma}=0$ which we know is wrong in
the presence of gauge fields . Noncommutative geometry , as it
will be obvious from equation (\ref{ginsparg-wilson}),  gives us
immediately the structure of  the chiral anomaly , indeed
(\ref{ginsparg-wilson}) can be rewritten in the form
\begin{eqnarray}
\bigg[{\Gamma}^R-{\Gamma}^L-\frac{2}{2l+1}D_{GF}\bigg]^2-4=\frac{2i}{(2l+1)^2}{\epsilon}_{ijk}{\sigma}_k(F_{ij}+F_{ij}^F)+\frac{8\sqrt{l(l+1)}}{(2l+1)^2}\bigg[{\phi}^F+\frac{\vec{A}^{F2}}{2\sqrt{l(l+1)}}\bigg],\label{ginsparg-wilson1}\nonumber\\
&&
\end{eqnarray}
where we have used extensively the identities
\begin{eqnarray}
&&D_F^2=(2l+1)^2-\frac{1}{4}(2l+1)^2({\Gamma}^R-{\Gamma}^L)^2\nonumber\\
&&D_{GF}^2=D_{GF}+({\cal
L}_i^F+{A}_i^F)^2+\frac{i}{2}{\epsilon}_{ijk}{\sigma}_kF_{ij}^F
.\nonumber
\end{eqnarray}
$F_{ij}$ in (\ref{ginsparg-wilson1}) denotes on the other hand the
{\it would-be} continuum curvature
\begin{eqnarray}
F_{ij}^F=F_{ij}+[A_i^F,A_j^F]~,~F_{ij}=[L_i,A_j^F]-[L_j,A_i^F]-i{\epsilon}_{ijk}A_k^F\nonumber
\end{eqnarray}
whereas ${\phi}^F$ denotes the fuzzy normal component of the gauge
field , namely ${\phi}^F=n_i^FA_i^F+A_i^Fn_i^F$.

Now if we multiply both sides of equation (\ref{ginsparg-wilson1})
by ${\psi}_F(a,A)$ from the right and ${\psi}_F(a,A)^{+}$ from the
left , we obtain the exact and elegant answer
\begin{eqnarray}
-tr({\Gamma}^R-{\Gamma}^L)+\frac{2l+1}{4}tr\bigg[\big[({\Gamma}^R-{\Gamma}^L)^2 - 4]\frac{1}{D_{GF}}\bigg]+\frac{1}{2l+1}trD_{GF}&=&\nonumber\\
\frac{i}{2(2l+1)}{\epsilon}_{ijk}tr\bigg[{\sigma}_k(F_{ij}+F_{ij}^F)\frac{1}{D_{GF}}\bigg]+\frac{2\sqrt{l(l+1)}}{2l+1}tr\bigg[\big[{\phi}^F+\frac{\vec{A}^{F2}}{2\sqrt{l(l+1)}}\big]\frac{1}{D_{GF}}\bigg].\label{ginsparg-wilson2}
\end{eqnarray}
The symbol $tr$ here denotes the {\it fuzzy} trace in the space
of fuzzy spinors , i.e $
tr(X)=\sum_{a}{\psi}_F^{+}(a,A)X{\psi}_F(a,A)$ . In particular the
cyclic property of the trace has to be used with care because of
the non-commutativity of the different ingredients , and
therefore the order in (\ref{ginsparg-wilson2}) is important.
This is more clear from the fact that the result of this
operation is still an element in the algebra ${\bf A}$ . The
limit is of course an ordinary spin trace .

We will now show that most contributions to the chiral anomaly
(\ref{anomaly}) are coming from high frequency modes of the
spectrum . First we have in the large $l$ limits ,
$\vec{n}^R=\vec{L}^R/{\sqrt{l(l+1)}}{\longrightarrow}\vec{n}$ ,
$\vec{n}^L=\vec{L}^L/{\sqrt{l(l+1)}}{\longrightarrow}\vec{n}$ and
hence the different chiralities must have the continuum limits
\begin{eqnarray}
&&{\Gamma}^R=\frac{1}{l+\frac{1}{2}}(-\vec{\sigma}.\vec{L}^R+\frac{1}{2}){\longrightarrow}-{\gamma}~,~{\Gamma}^L=\frac{1}{l+\frac{1}{2}}(\vec{\sigma}.\vec{L}^L+\frac{1}{2}){\longrightarrow}{\gamma}.\label{limit}
\end{eqnarray}
But from the free fuzzy eigenvalues equation
\begin{eqnarray}
\frac{D_F}{2l+1}{\psi}_F(a,0)=\frac{{\Gamma}^R+{\Gamma}^L}{2}{\psi}_F(a,0),
\end{eqnarray}
we can easily see that for all the infrared modes $a<<2l$ , the
above limits (\ref{limit}) are indeed satisfied since
\begin{eqnarray}
Lim_{l{\longrightarrow}{\infty}}\big[\frac{D_F}{2l+1}{\psi}(a,0)\big]=0.
\end{eqnarray}
But for ultraviolet modes $a{\sim}2l$ we have instead
\begin{eqnarray}
Lim_{l{\longrightarrow}{\infty}}\big[\frac{D_F}{2l+1}{\psi}(a,0)\big]={\pm}1.\label{364}
\end{eqnarray}
The sign is $+1$ if $e_a$ is a positive energy eigenvalue
and $-1$ if $e_a$ is a negative energy eigenvalue . This
means in particular that the limits (\ref{limit}) are not valid in
the UV domain and therefore the statement $
tr\big[{\Gamma}^R-{\Gamma}^L\big]=0 $ itself which is a
consequence of (\ref{importantfact}) is in fact a statement about
the IR modes only of the theory .

Equation (\ref{364}) also means that in the limit the operator
$\frac{D_F}{2l+1}$ restricted to the above UV modes is acting as
the sign operator $F_F=\frac{D_F}{|D_F|}$ and is not identically
zero as we would have expected . So if we are only restricted to
the high frequency part of the spectrum we can, for all practical
purposes , set in the large $l$ limit
\begin{eqnarray}
\frac{D_F}{2l+1}|_{a{\sim}2l}{\longleftrightarrow}{F}_F=-i{\Gamma}^{'}{F}_{w}.\label{crucial1}
\end{eqnarray}
In the continuum limit the above equation reduces to the identity
$ {\cal F}=-i{\gamma}{\cal F}_{w}$ where ${\cal F}$ is the sign
of the Dirac operator ${\cal D}$  , while ${\cal F}_{w}$ is the
sign of the Watamura Dirac operator ${\cal D}_w$ , i.e ${\cal
F}_{w}=\frac{D_{w}}{|D_{w}|}$ \cite{watamuras}. ${\Gamma}^{'}$ on
the other hand is another chirality operator which is different
from both ${\Gamma}^R$ and ${\Gamma}^L$ , but has the same
continuum limit and that is all we will need . It was however
constructed explicitly in \cite{ydri} . The fuzzy version of the
Watamuras Dirac operator $D_{w}$ annihilates the top modes
$j=2l+\frac{1}{2}$ and hence a proper regularization of
$F_{w}=\frac{D_{w}}{|D_{w}|}$ is required . We made  therefore the
natural replacement
\begin{eqnarray}
{F}_{w}{\longrightarrow}F_{\Lambda
w}&=&\frac{D_{w}}{|D_{w}|}~{\rm for}~{\rm
all}~j<2l+\frac{1}{2}\nonumber\\
&=&+1~{\rm for}~j=2l+\frac{1}{2}.\label{regularization}
\end{eqnarray}
As $F_{w}$ is a sign operator  , the only other completely
equivalent choice will be $F_{\Lambda w}=-1$ on the modes
$j=2l+\frac{1}{2}$ .

This funny behaviour in the UV is the source of the anomaly which
is essentially a quantum mechanical effect as of the presence of
the full propagator $\frac{1}{|D_{GF}|}$ in
(\ref{ginsparg-wilson2}) . Now we expect that the behaviour of
the IR modes will not change as compared to the free case and
therefore all the anomaly will be captured by the UV modes .
Indeed for any field configuration we will have in the continuum
limit the identity
\begin{eqnarray}
&&-\frac{1}{2(2l+1)}Tr{\lambda}^Ltr_{a<<2l}({\Gamma}^R-{\Gamma}^L)=-\frac{1}{2}\int_{{\bf S}^2} \frac{d{\Omega}}{4{\pi}}{\lambda}(\Omega)tr_{a<<2l}(-2{\gamma})=0.\nonumber\\
\end{eqnarray}
We know that the trace is $0$ because we know that all the
infrared gauge-invariant modes are paired .

The leading contribution to the chiral anomlay will be given in
the large $l$ limit by
\begin{eqnarray}
\frac{i}{4(2l+1)^2}{\epsilon}_{ijk}Tr{\lambda}^Ltr\bigg[{\sigma}_k(F_{ij}+F_{ij}^F)\frac{1}{D_{GF}}\bigg]
&{\simeq}&\frac{i}{2(2l+1)}{\epsilon}_{ijk}Tr{\lambda}^Ltr_{a{\sim}2l}\bigg[{\sigma}_kF_{ij}\frac{1}{|D_{F}|^2}F_{F}\bigg],\label{447}\nonumber\\
\end{eqnarray}
where the approximation ${\simeq}$ clearly becomes exact in the
strict large $l$ limit, i.e all ignored terms in (\ref{447}) are
subleading and vanishes identically in this strict limit . We
have also used the fact that $F_{ij}^F$ for large $l$ is
essentially $F_{ij}$, and that the gauge field can always be
treated as a weak perturbation in the continuum limit in the
sense that $<\vec{n},l|A_i^F|\vec{n},l><<l$ .

By using the crucial result (\ref{crucial1}) we can rewrite the
above axial anomaly as
\begin{eqnarray}
\frac{i}{4(2l+1)^2}{\epsilon}_{ijk}Tr{\lambda}^Ltr\bigg[{\sigma}_k(F_{ij}+F_{ij}^F)\frac{1}{D_{GF}}\bigg]
&{\simeq}&\frac{1}{2(2l+1)}{\epsilon}_{ijk}Tr{\lambda}^Ltr_{j=2l+\frac{1}{2}}\bigg[{\sigma}_kF_{ij}\frac{1}{|D_F|^2}{\Gamma}^{'}\bigg]\nonumber\\
&+&\frac{1}{2(2l+1)}{\epsilon}_{ijk}Tr{\lambda}^Ltr_{a{\sim}2l}\bigg[{\sigma}_kF_{ij}\frac{1}{|D_F|^2}{\Gamma}^{'}F_{{\Lambda}w}\bigg],\nonumber\\
\end{eqnarray}
where we have separated the top modes $j=2l+\frac{1}{2}$ from the
rest of the UV's as they are not paired to anything else which is
clear from equation (\ref{regularization}). It is this same
property which allows us to conclude that the second term above
vanishes identically and we are left with
\begin{eqnarray}
\frac{i}{4(2l+1)^2}{\epsilon}_{ijk}Tr{\lambda}^Ltr\bigg[{\sigma}_k(F_{ij}+F_{ij}^F)\frac{1}{D_{GF}}\bigg]
&{\simeq}&\frac{1}{2(2l+1)}{\epsilon}_{ijk}Tr{\lambda}^Ltr_{j=2l+\frac{1}{2}}\bigg[{\Gamma}^{'}{\sigma}_kF_{ij}\frac{1}{|D_F|^2}\bigg],\nonumber\\
\end{eqnarray}
where we have now used the fact that in the large $l$ limit the
trace $tr$ is almost cyclic and therefore all corrections  are
subleading . Since we are analyzing the continuum limit and since
${\Gamma}^{'}$ has the canonical continuum limit ${\gamma}$ we
can set ${\Gamma}^{'}=\vec{\sigma}.\vec{n}^L$ without any loss of
generality . The actual expression is of course much more
complicated as was shown in \cite{ydri} . We can then use the
identity
\begin{eqnarray}
{\epsilon}_{ijk}{\Gamma}^{'}{\sigma}_kF_{ij}={\epsilon}_{ijk}n_k^LF_{ij}-2i{\sigma}_jn_i^LF_{ij}.
\end{eqnarray}
The last term above is identically zero in the strict
$l{\longrightarrow}{\infty}$ limit provided the gauge field
${A}_i$ is projected appropriately onto the sphere , i.e we have
$n_iF_{ij}=0$ and we end up with the final result
\begin{eqnarray}
\frac{i}{4(2l+1)^2}{\epsilon}_{ijk}Tr{\lambda}^Ltr\bigg[{\sigma}_k(F_{ij}+F_{ij}^F)\frac{1}{D_{GF}}\bigg]&{\simeq}&\frac{1}{2(2l+1)}{\epsilon}_{ijk}Tr{\lambda}^Ltr_{j=2l+\frac{1}{2}}\bigg[n_k^LF_{ij}\frac{1}{|D_F|^2}\bigg]\nonumber\\
&{\simeq}&{\epsilon}_{ijk}Tr{\lambda}^Ln_k^LF_{ij}\frac{1}{|D_F|^2}.\nonumber\\
\end{eqnarray}
We have now used that $\frac{1}{|D_F|^2}$ will behave exactly as a
volume form , i.e it does not contain spin indices and hence
$n_k^LF_{ij}\frac{1}{|D_F|^2}$ acts as an identity in the space
of spinors . This identity is exactly
$2(2l+\frac{1}{2})+1=2(2l+1)-$dimensional in the top mode sector
$j=2l+\frac{1}{2}$ .

The last step is to convert the above result into an integral .
To this end we use the generalized coherent states and the answer
is
\begin{eqnarray}
S_{\theta
F}{\longrightarrow}S_{\theta}={\epsilon}_{ijk}Tr{\lambda}^Ln_k^LF_{ij}\frac{1}{|D_F|^2}=\frac{2lnl}{(4{\pi})^5}\int_{{\bf
S}^2}\frac{d{\Omega}}{4{\pi}}{\lambda}(\vec{n}){\epsilon}_{ijk}n_kF_{ij}(\vec{n}).\label{beautiful}
\end{eqnarray}
The last thing to do is to verify that the high-frequency
contribution of the other terms in (\ref{ginsparg-wilson2}) are
vanishing in the continuum limit . This is indeed a trivial
exercise to do so we skip here the detail .

Putting together this last result (\ref{beautiful}) with the
result (\ref{anomaly0}) we obtain exactly the local chiral
anomaly equation
\begin{eqnarray}
{\cal
L}_i(\bar{\chi}{\sigma}_i{\gamma}{\chi})(\vec{n})={\epsilon}_{ijk}n_kF_{ij}.
\end{eqnarray}

\bibliographystyle{unsrt}

\end{document}